\documentclass[fleqn,usenatbib]{mnras}

\usepackage{newtxtext,newtxmath}
\usepackage{xcolor}
\usepackage[T1]{fontenc}

\DeclareRobustCommand{\VAN}[3]{#2}
\let\VANthebibliography\thebibliography
\def\thebibliography{\DeclareRobustCommand{\VAN}[3]{##3}\VANthebibliography}

\usepackage{soul}
\renewcommand{\hl}[1]{#1}
\usepackage{graphicx}	
\usepackage{amsmath}	

\title[Star Cluster Formation with External Driving]{Star Cluster Formation in Clouds with Externally Driven Turbulence.}

\author[J. D. Smith et al.]{
Jamie D. Smith,$^{1}$\thanks{E-mail: j.smith49@herts.ac.uk (JDS)}
James E. Dale,$^{1}$
Sarah E. Jaffa$^{1,2}$
and Martin G. H. Krause$^{1}$
\\
$^{1}$Centre for Astrophysics Research, Department of Physics, Astronomy and Mathematics, University of Hertfordshire, College Lane, Hatfield, Hertfordshire AL109AB, UK\\
$^{2}$Advanced Research Computing, University College London, London, WC1H 9NE, UK
}

\date{Accepted XXX. Received YYY; in original form ZZZ}

\pubyear{2021}

\begin{document}
\label{firstpage}
\pagerange{\pageref{firstpage}--\pageref{lastpage}}
\maketitle
\begin{abstract}
Star clusters are known to be formed in turbulent molecular clouds. How turbulence is driven in molecular clouds and what effect this has on star formation is still unclear. We compare a simulation setup with turbulent driving everywhere in a periodic box with a setup where turbulence is only driven around the outside of the box. We analyse the resulting gas distribution, kinematics, and the population of stars that are formed from the cloud. Both setups successfully produce a turbulent velocity field with a power law structure function, the externally driven cloud has a more central, monolithic, clump, while the fully driven cloud has many smaller, more dispersed, clumps. The star formation follows the cloud morphology producing large clusters, with high star forming efficiency in the externally driven simulations and sparse individual star formation with much lower star formation efficiency in the fully driven case. We conclude that the externally driven method, which resembles a Global Hierarchical Collapse (GHC) scenario, produces star clusters that more closely match with observations.
\end{abstract}




\section{Introduction}

Star clusters (SC) are formed in molecular clouds (MC), \citep[e.g.][for reviews]{ladaladaclusterincloud,krausereview} that are observed to be turbulent \citep{larson,elmegreenscalo04}, have filamentary structures \citep[e.g.][]{recentfilamentUCHII,filametsOrion,obsFilaments}, and can inherit the turbulent characteristics of the surrounding medium \citep[][]{2014prplDobbsinherit}.

The driving source behind turbulence on the galactic scale is a debated subject \hl{and many studies have attempted to distinguish between the different driving modes} \citep[\hl{e.g.}][]{Agertz2016,Faucher2013,Goldbaum2016,Krumholz2016,Krumholz2018}. Some studies credit galactic dynamics, mainly the interaction between the ISM and the spiral arms \citep[e.g.][]{drivingsource1}. Others credit the feedback that results from star formation for providing an energy source for the turbulence, mainly supernovae \citep[e.g.][]{2005AvillezBreitschwerdt,superdriven}.

\cite{molturbfromSNe} study whether external supernovae are sufficient to sustain turbulence in molecular clouds and find supernovae sufficient at early stages but require support from other sources later in the cloud's development.

Other studies credit global hierarchical collapse (GHC) for the source of the internal motions in molecular clouds \citep[e.g.][]{enriqueGroupsGHC,GHC_enrique,2020ISMGHC,2020MNRASphotoionizingGHC}. This scenario describes a hierarchy of scales with each scale accreting matter from larger scales.
Colliding flow simulations demonstrate an idealised representation of the GHC scenario \citep[e.g.][]{collisionDobbs2,collisionDObbs,collidingmhdwurster}. This involves two streams of gas which collide with large velocity creating regions of higher density which can then continue to collapse.

Observations of the ISM, from the $\textit{Herschel Space Observatory}$, and simulations show that interstellar clouds are host to intricate filamentary structures \citep[e.g.][]{Andr__2014,filaments,obsFilaments}. Such filaments can be explained in both scenarios \citep[e.g.][]{krausereview}: in the GHC picture, the filaments form by anisotropic collapse, whereas for a globally turbulently supported ISM, one expects filaments from shock waves \citep[\hl{e.g.}][]{Burkhart2019,federrath2016,mocz2017}. Therefore more details are needed to distinguish between the scenarios.

By definition the GHC scenario is dominated by large-scale, external driving, \hl{either from large-scale gravitational collapse of a cloud that has exceeded its Jeans limit by gas accumulation, or actively colliding flows}. The bulk kinetic energy is converted into heat and turbulence when the flows collide from different directions. The alternative is to assume that the cloud is supported by turbulence due to sources that include the aforementioned ones, and collapse \hl{occurs locally in filaments and more slowly elsewhere due to this support} (gravoturbulent scenario). \hl{In detail, the two scenarios have some properties that may be hard to distinguish observationally} (compare discussion in \cite{krausereview}). \hl{Both predict turbulence at various scales. While the gravoturbulent scenario identifies a scale below which gravitational collapse dominates, GHC has various scales, possibly also the largest one where collapse dominates. A naive expectation might therefore be that GHC promotes more clustering on larger scales compared to a gravoturbulent setting.}

\hl{Some authors have investigated the effect of the driving scheme on simulations and generally, a strong influence is found. For example,} \cite{Offner2008decaying} \hl{found that decaying turbulence produces higher-multiplicity stellar systems.} \cite{Girichidis2011} \hl{compared solenoidal with compressional driving and found a strong influence on the formation of clusters.} \cite{Lane2022lesswrong} \hl{compare spherical cloud simulations to periodic box simulations with and without external driving. They found that star formation is considerably slower in box simulations, highlighting the importance of a global collapse mode for star formation.}

\hl{In this work we investigate a combination of the previously studied conditions, with a particular focus on comparing the GHC and the gravoturbulent scenario with setups that are otherwise very similar.} Using a series of simulations that allow for a direct comparison of the scenarios. \hl{In particular, we compare a setup that starts from equilibrium turbulence, designed to implement the gravoturbulent scenario with energy input everywhere at large scales, to a setup that inputs energy externally at the edge of the box in order to better numerically represent what we see in nature. This is also reminiscent of clouds in larger scale simulations} \citep[e.g.][]{Duarte2016,Seifried2017,molturbfromSNe,Pettitt2020},
\hl{and is designed to implement the property of the GHC scenario that there is some region and length-scale range without driving, such that gravitational collapse can proceed more freely on such scales. Indeed, we find significant differences in the clustering properties of these simulations.}

In Section \ref{method} we describe the relevant details of the Smooth Particle Hydrodynamic simulations that we perform and the analysis that we do on them. In Section \ref{sec:results} we present the main outputs of the simulations and the products of the analysis, we show that our driving methods successfully drive turbulence and create star clusters, though with interesting differences. In Section \ref{sec:discussion} we discuss our results and suggest explanations for any observed differences. Finally in Section \ref{sec:conclusions} we summarise this work and conclude that external driving can result in more realistic star clusters than an equilibrium turbulence setup.

\section{Method}
\label{method}

For all simulations in this work we use the Smoothed Particle Hydrodynamics code PHANTOM \citep{Phantomsph}. We use this code to simulate a $20\,\mathrm{pc}$ periodic box with an isothermal equation of state with a temperature of $10\,\mathrm{K}$, a sound speed of $200\,\mathrm{m\,s}^{-1}$, and polytropic index of one.

Turbulence is driven using Ornstein-Uhlenbeck stochastic driving, to apply an acceleration field to the particles depending on their location within the box. By default this is done everywhere throughout the lifetime of the simulation but we modify this for some simulations as described below. The 'strength' of the driving is moderated by an amplitude factor, which we adjust to obtain an equilibrium mach number in some of the simulations.

Self-gravity is treated by separating the effect into short range and long range components. The short range is treated using a summation of the contributions from individual nearby particles, and the long range uses a kd-tree to hierarchically group the distant particles and compute the contribution from each group.
We use this simulation code to create turbulent boxes with various properties, varying the initial uniform density, and thus the total mass in the box, the region that the turbulent driving is applied to, and the strength of the turbulent driving if needed.
Sink particles \citep{sinkparticleBate1995} are used to represent stars or small multi-star systems. They form when the density exceeds a given critical density allowing the simulation to track the evolution below fragmentation and they are evolved on shorter timesteps than the SPH particles. The sink particles are allowed to accrete gas, and interact with their surroundings only through gravity.

For our numerical experiments we use two, initially uniform, densities which results in total masses of 5,000 and 10,000 \hl{solar} masses (labelled '$5\mathrm{k}$' and '$10\mathrm{k}$' respectively, more simulations parameters in Table \ref{tab:simulations}).  This box is then driven for a simulated time of $21\,\mathrm{Myr}$ to ensure that the turbulence is in statistical equilibrium before self-gravity is switched on, and continues to be driven for the duration of the simulation. Sink particles are formed at a critical density such that the mass of 100 particles exceeds the Jeans mass following the resolution criteria from \citep{sphresolustion}. The critical value for the '$5\,k$' simulations is $2.22\times 10^{-17}\,\mathrm{g\,cm}^{-3}$ and for the '$10\,k$' simulations is $5.56\times 10^{-18}\,\mathrm{g\,cm}^{-3}$.

The two main setups we compare in this work are the fully driven box and the externally driven box. The fully driven runs represent the Gravoturbulent scenario where we drive turbulence \hl{at all locations in the box at scales corresponding to wavenumbers between $2\pi$ and $6\pi$}. The externally driven runs are motivated for example by the work of \cite{molturbfromSNe}, and share some properties of the global hierarchical collapse scenario.

To simulate the external energy input for the turbulent driving, we restrict the driving to a region of $2\,\mathrm{pc}$ at each edge of the box leaving a $16\,\mathrm{pc}$ per side box that is not driven. We do this by simply setting the accelerations due to turbulence to zero inside the internal box. This velocity field of this internal box is then only turbulent due to energy cascade from the driven external region. This also produces a net inwards motion into the central region which produces a similar effect to that of a 'colliding flow' simulation \citep[e.g.][]{collisionDobbs2,collisionDObbs}.

Additional simulations are performed to check a variety of different situations. First, in order to check whether any result is due to the amount of energy being introduced rather than the manner of introduction, a fully driven box is used with much weaker driving than the main simulations. To achieve this the stirring amplitude is reduced so that the internal energy in the box is comparable to the externally driven simulations. Additionally, repeats of the main simulations are performed to check for consistency - the same setups are used with identical parameters chosen, only the random turbulent driving changes.
The simulations are named based on the number of particles, ($2m$ meaning two million), the total mass in the box in solar masses ($5k$ or $10k$), and an 'e' denoting externally driven. Simulations that are repeated are given a 'mk2' suffix. An overview of the simulations is given in \ref{tab:simulations}.

\subsection{Analysis Techniques}
\label{analysis}
\subsubsection{Structure Functions}
\label{sub:strucfunc}
We use velocity structure functions of second order to diagnose a turbulent velocity field by comparing the structure function to the expected power law. Velocity structure functions are commonly used alongside, or instead of, the velocity power spectrum \citep[e.g.][]{StrucFunc,gridstructures,howdoStrucFunc}. We calculate the structure functions by taking a sub-sample of the particles, we then compare every 'sample' particle with every particle from the simulation recording the difference in the velocities and the distance between the particles. The distances are then binned and the square of the average of the velocity differences from each bin are taken. For consistency between the fully driven and externally driven cases we will only consider the internal $16\,\mathrm{pc}$ box for this calculation in both cases.
\begin{equation}
    \label{structure func}
    S(\mathrm{d}r) = \langle (\mathbf{v}_i - \mathbf{v}_j)^2\rangle_{\mathrm{bin}} = \langle \delta v^2 \rangle _{\mathrm{bin}}
\end{equation}

\subsubsection{Initial Mass Function}
\label{sub:MassFunc}
We use the mass function of the sink particles to determine whether we produce a realistic distribution of masses. We expect our mass functions to show a power law at the high mass end, but we note that the location of the peak of our mass functions will depend on the mass resolution of the simulation as well as on the physics we included and therefore may not agree with observations of the IMF.

The nature of the IMF of a stellar population is connected to the turbulent structure of the cloud the population formed within. \cite{effectofnonIMFslope} shows that the slope of the velocity power spectrum of a cloud is linked to the slope of the high mass end of the initial mass function for that cloud, and that a shallower power spectrum leads to a shallower IMF and therefore more high-mass stars.

\subsubsection{Cluster Morphology}
\label{sub: cluster morph}
We use a friends of friends algorithm \citep{davis1985Groups} which is often used  to group particles in simulations according to some distance parameter \citep[e.g.][]{enriqueGroupsGHC}, any particle or existing group within that distance is then considered to be part of the same group. We perform this at the end of the simulation and we do not track the groups development over time. The distance parameter we use was found by trial and error until a small change didn't drastically change the groupings. 
We also look at the mass to radial distance relation. To do this we calculate the centre of mass for the group and then plot each member of the groups mass versus its radial distance.

\subsubsection{Star Formation Efficiency}
We calculate the Star Formation Efficiency (SFE) by taking the current number of particles that have been accreted onto sink particles and dividing by the total number of SPH particles. As all SPH particles have uniform mass this gives us the percentage of initial mass that has formed stars.

\begin{equation}
\label{eq:SFE}
    SFE = \frac{\sum N_\mathrm{SPH,acc}}{N_\mathrm{SPH,tot}}
\end{equation}

\begin{table*}
\centering
\caption{Table showing various simulation details.}
\label{tab:simulations}
\begin{tabular}{|l|l|l|l|l|l|l|l|l|} 
\hline
Simulation & $N_\mathrm{particles}$ & Particle Mass~ & Turbulence & Duration (Myr) & $M_\mathrm{total}$ & $N_\mathrm{sinks}$        & Min Sink mass & Mach  \\ 
\hline
2m5k       & 2,000,000              & 0.0025         & Full       & 42.2           & 5,000              & \hl{9}   & 0.25          & \hl{25.1}  \\ 
\hline
2m5ke      & 2,000,000              & 0.0025         & External   & 26.8           & 5,000              & \hl{316} & 0.25          & \hl{15.7}  \\ 
\hline
2m10k      & 2,000,000              & 0.005          & Full       & 42.2           & 10,000             & \hl{217} & 0.5           & \hl{25.0}  \\ 
\hline
2m10ke     & 2,000,000              & 0.005          & External   & 24.6           & 10,000             & \hl{420} & 0.5           & \hl{13.1}  \\
\hline
\end{tabular}
\end{table*}

\section{Results and Analysis}
\label{sec:results}

\subsection{Large-scale gas distributions}
\label{sec:morphology}
\begin{figure*}
	
    \includegraphics[width=0.8\textwidth]{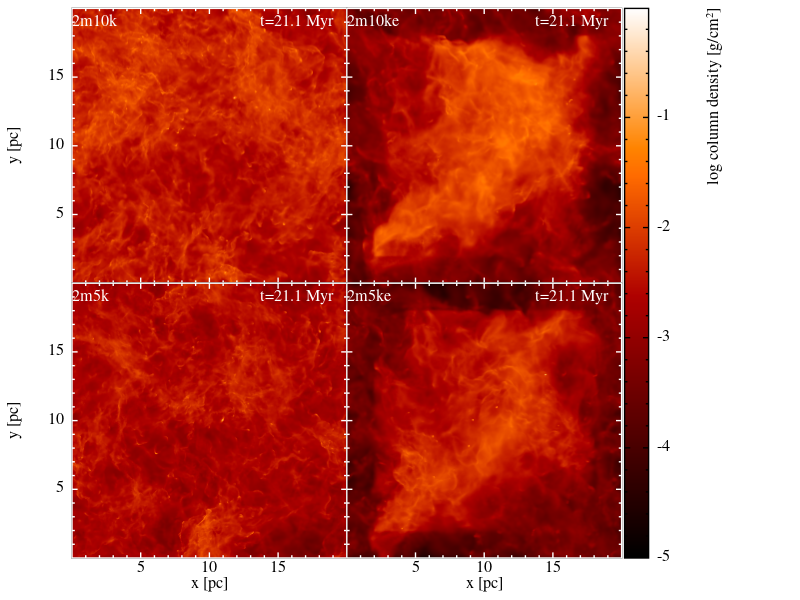}
	\includegraphics[width=0.8\textwidth]{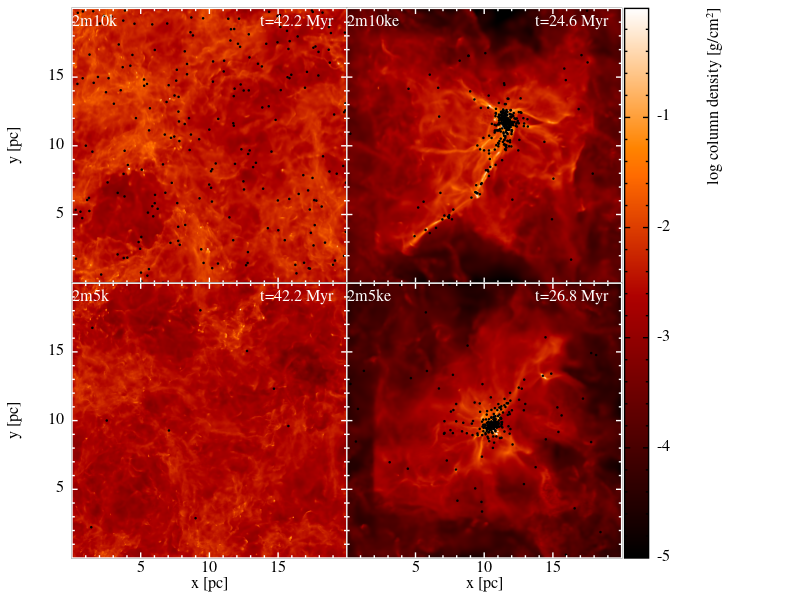}
    \caption{Logarithmic column density plots of the state of the simulations at two different timesteps: \textbf{Upper} just before gravity is switched on ($21\,\mathrm{Myr}$). \textbf{Lower} the final timestep from each individual simulation which is labelled on each panel. Sink particles are indicated by black dots.}
    \label{fig:pre-grav}
\end{figure*}

Figure \ref{fig:pre-grav} shows column density maps for our four main simulations. In every panel, Simulations with external driving are on the right, and ones with full driving on the left. The high-mass clouds are in the upper parts and the low-mass ones are in the lower parts of the panels. External driving consistently produces a central dense clump, or sequence of clumps, connected and surrounded by filamentary structure. In the fully driven box we do not observe any noticeable central hub system, instead we see many small clumps spread around the box without any significant filaments.

\textit{\textit{}}

\subsection{Gas Kinematics}
\label{sec:kinematics}

\begin{figure*}

	\includegraphics[width=0.45\textwidth]{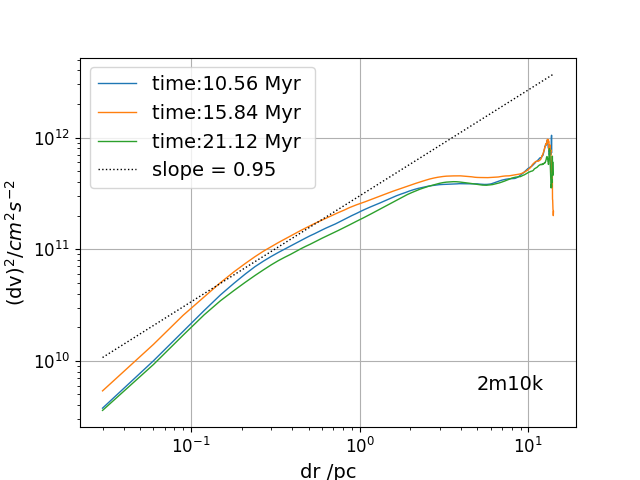}
	\includegraphics[width=0.45\textwidth]{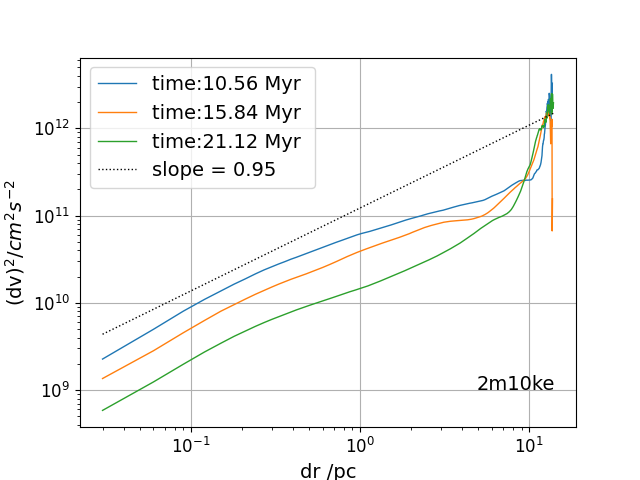}
	\includegraphics[width=0.45\textwidth]{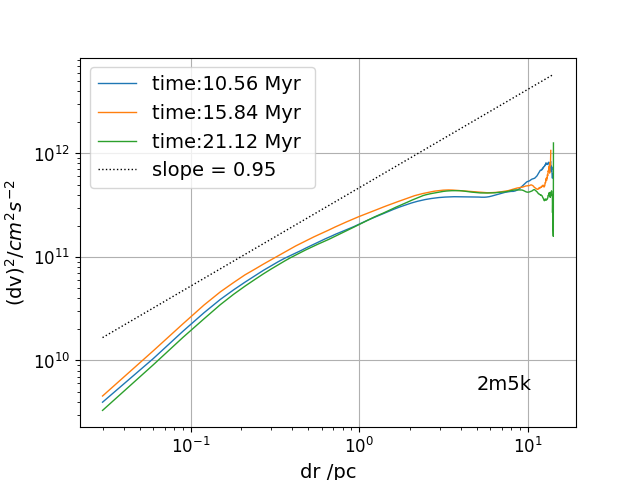}
	\includegraphics[width=0.45\textwidth]{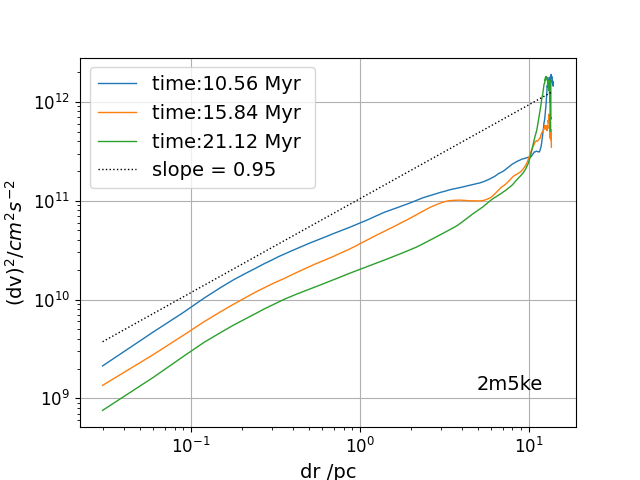}
    \caption{Structure functions for the simulations taken just before self-gravity is switched on, plotted with solid lines. \hl{A power-law with the expected slope of 0.95 for compressible, supersonic turbulence} \citep{kowal2007,kritsuk2007} \hl{is plotted as a dotted line}. Left panels: show the structures for the externally driven velocity fields. Right panel: shows the structures for the fully driven simulation.}
    \label{fig:structures}
\end{figure*}

We show the structure functions of all simulations in Fig. \ref{fig:structures}. The external driving of the velocity field is sufficient to produce kinematics in the interior of the box.
Both the fully driven and the externally driven velocity structure functions show a power law in the intermediate scales \hl{with an index comparable to 0.95, which has been found from high resolution simulations of compressible turbulence} \citep{kowal2007,kritsuk2007}. \hl{This result is similar to comparable simulations of }\cite{Lane2022lesswrong} \hl{who use a quasi-Lagrangian mesh-less finite-mass method}. At scales below $0.2\,\mathrm{pc}$ the structures break due to the artificial viscosity with the break in the fully driven simulations being somewhat more severe than in the externally driven ones. We surmise that the severity of the break is greater in the fully driven case as the energy levels are greater than in the external case.
The shape of the structure functions have converged at $21\,\mathrm{Myr}$, meaning that the turbulent velocity field has reached an equilibrium state. For the fully driven simulations this state is at constant kinetic energy whereas the externally driven simulations have a constantly declining energy, i.e., the turbulence decays at a steady rate.

\subsection{Stellar Population}
\label{sec:StellarPop}
\begin{figure*}
	\includegraphics[width=0.45\textwidth]{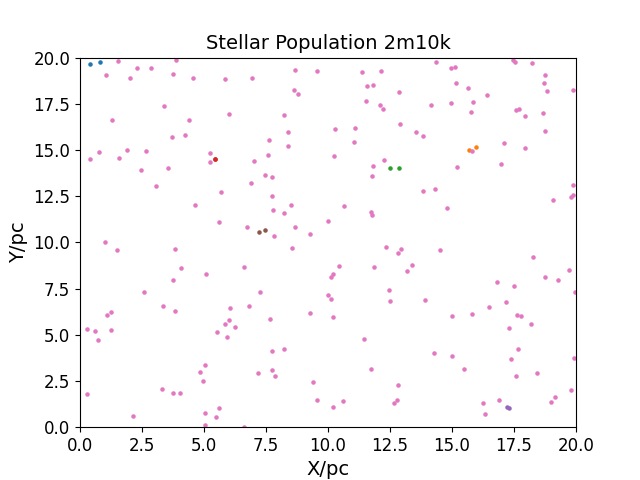}
	\includegraphics[width=0.45\textwidth]{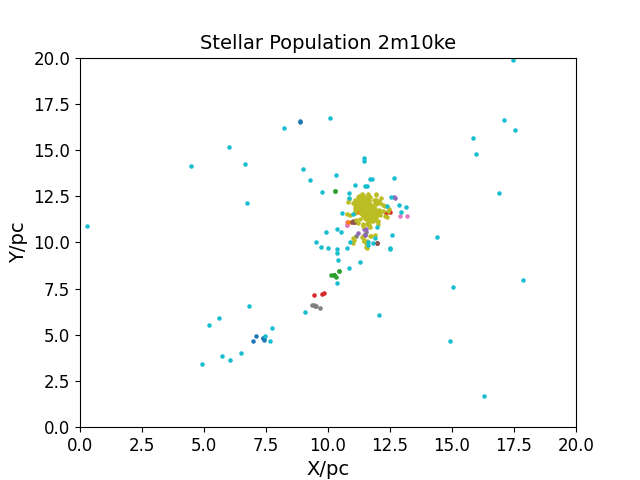}
	\includegraphics[width=0.45\textwidth]{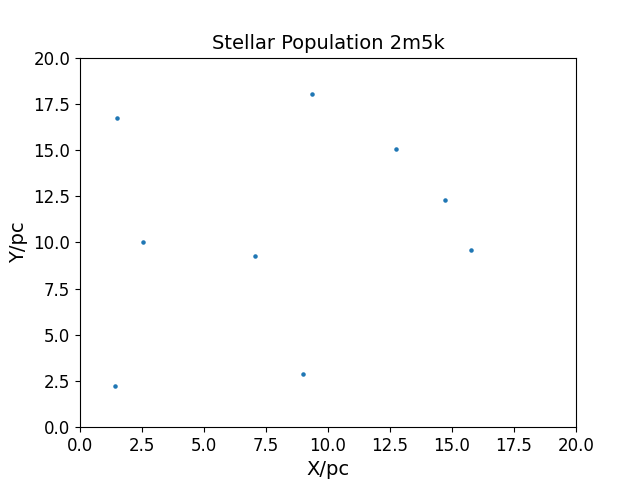}
	\includegraphics[width=0.45\textwidth]{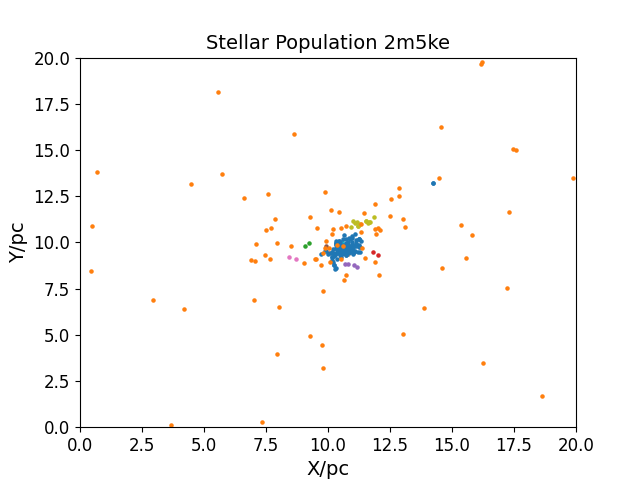}
    \caption{Plots showing the grouping of the stellar populations taken at the end of each simulation. The fully driven simulations to the left, external to the right. The '10k' simulations at the top with the '5k' below.}
    \label{fig:groups}
\end{figure*}

The morphology of the stellar populations that are formed within our simulations (Fig \ref{fig:groups}) naturally follow the gas morphology described in section \ref{sec:morphology}. The fully driven boxes show dispersed sink particle formation, with individual or pairs of sink particles spread across the box. The externally driven boxes shows large clusters of star formation within the clump systems described in \ref{sec:morphology}, resulting in more sink particles forming in groups at the centre of the dense structures.

\begin{figure}
	\includegraphics[width=0.47\textwidth]{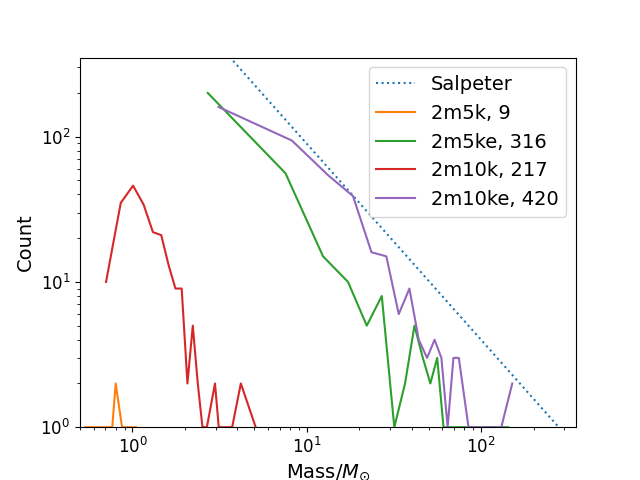}
    \caption{The \hl{sink} mass functions for each of the four main simulations are shown and labelled the number of sinks given for each simulation. The Salpeter power law is also plotted for reference.}
    \label{fig:massfuncs}
\end{figure}
\begin{figure}
    \centering
    \includegraphics[width=0.47\textwidth]{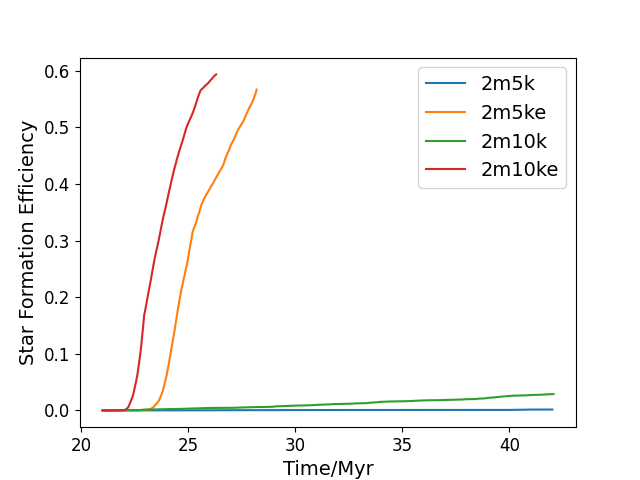}
    \caption{Plot with the star formation efficiencies for each simulation labelled.}
    \label{fig:SFE}
\end{figure}

\begin{figure}
	\includegraphics[width=0.45\textwidth]{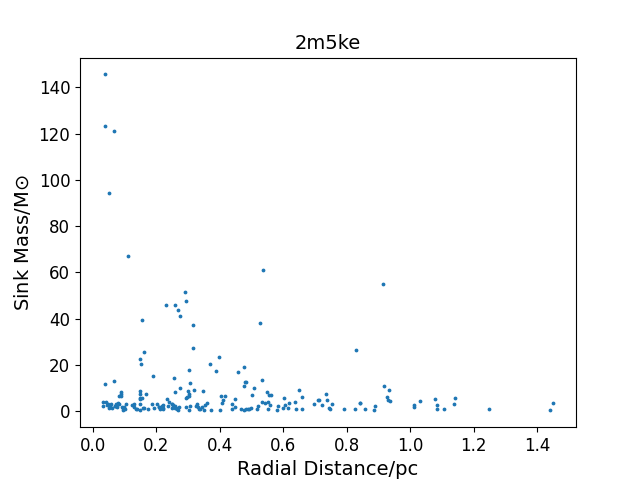}
	\includegraphics[width=0.45\textwidth]{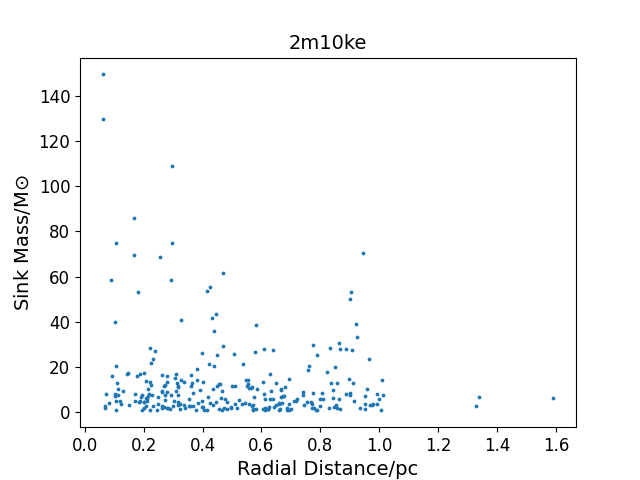}
    \caption{The radial mass distribution of the largest group from each of the externally driven simulations}
    \label{fig:radialmass}
\end{figure}

The mass functions (fig. \ref{fig:massfuncs}) from the external simulations show a much larger stellar population than the fully driven simulations.The fully driven run with $5,000$ solar masses (2m5k) does not produce enough stars to produce an informative mass function. The externally driven mass functions peak at higher mass than the fully driven mass functions.

Figure \ref{fig:SFE} shows the Star Formation Efficiency (SFE) from each of the main runs. In both of the fully driven cases star formation starts very soon after self-gravity is switched on and continues at a slow rate resulting in a SFE of $<5\%$ in both cases. The externally driven cases begin forming stars slightly later but at a much greater rate resulting in a SFE of between $30\%$ and $50\%$ within one to a few Myr. \hl{We believe that the low SFE found in our fully driven simulations can be attributed to low mass resolution relative to previous work in the literature.}
\section{Discussion}

\label{sec:discussion}

The two different driving methods produce simulations with some similarities and some stark differences. Both methods produce velocity structure functions comparable to the expectations for supersonic turbulence as displayed in figure \ref{fig:structures}. The fully driven turbulence producing equilibrium turbulence and the externally driven producing constantly decaying turbulence. External driving leads to the formation of a large filamentary clump structure, similar to what is found in colliding flow simulations \citep[e.g.][]{collisionDobbs2,collisionDObbs,collidingmhdwurster} whereas the full driving formed smaller, more dispersed clumps. The mass function of the externally driven clusters show a reasonable power law with a slightly steeper slope than the Salpeter slope. This suggests that we do not form as many massive stars as we would expect. The fully driven clusters don't produce enough stars to produce an informative mass function.

We performed supplementary simulations to check for potential other causes for the differences we observed. We used a fully driven simulation with much weaker turbulent driving to test whether the results were due to the strength of the turbulence rather than the driving method. The fully, but weakly, driven box displays similar results to the regular fully driven simulations and therefore the strength of the turbulence is not alone responsible for the observed results. Duplicate runs of the main simulations were also performed to check for consistency and produce similar results \citep{Jaffa2022}.

We also note that our simulations differ in rms Mach number, higher mach numbers are observed to accelerate star formation \citep{motta2016}. We see lower star formation in our fully driven simulations that have higher star formation, this emphasises that the difference is in the driving method.

\cite{FeedbackStopsFormation} simulates a number of collapsing clouds with varying initial virial parameter. Their pre-feedback snapshots show similar morphology to our externally driven clouds with central clumps and sink particles concentrated around the clumps and filaments.

The Stellar distributions from the externally driven simulations appears consistent with examples of star clusters from GHC simulations \citep[e.g.][]{GHC_enrique}. The fully driven simulations produce disperse stellar populations with no notable groups. This is similar to the results from \citep[][]{gravoTEOS} who show that a polytropic index of unity, or higher, leads to isolated star formation with clustered formation resulting from lower polytropic index. The polytropic index of our simulations is 1. \hl{Our results are in good agreement with the ones of} \cite{Lane2022lesswrong} \hl{who compare a periodic-box setup with a spherical-cloud one. They also find more star formation for the cloud setup, concluding on the importance of a global collapse mode for star cluster formation. We confirm their result with our somewhat different setup.}

The stellar IMF, which describes the mass distribution of stars in a stellar population, is an important metric across many scales in astrophysics. The IMF is particularly useful as it compares well between simulations and observations, although there are potential limitations with observations being limited due to their sensitivity where simulations are not.

Neither of the fully driven simulations produce enough sink particles to create an informative mass function, beyond an approximation of the peak mass. The externally driven simulations do produce enough sink particles to be informative. The peaks of the externally driven mass functions are located at higher mass than the fully driven ones. This could be because of material being fed into the central hub sustaining accretion, whereas, the sinks in the fully driven simulation are starved of gas shortly after they form.

The externally driven simulations, while producing a reasonable number of sink particles, don't fully sample the high mass part of the IMF. This results in a slightly steeper power law than we would expect to see. Due to the mass resolution limitations of the simulations we do not expect to accurately reproduce the low mass end of the IMF.

The difference in the slopes of the SFEs, and in their final SFE, for the different cases we attribute to the driving methods. The external driving results in a net inwards motion of the gas resulting in a large clump and filament system which provides a supply of gas for stars to form and continue to accrete. The full driving does not create this central concentration of gas and therefore less star formation occurs and there is less sustained supply of gas for the stars to accrete. The high SFE range for the externally driven cases is compatible with expectations from star clusters formed in colliding flows, even when considering ionisation feedback \citep[][]{2014EnriqueZamora}. Our simulations have no feedback prescription which has been shown to limit star formation \citep[e.g.][]{FeedbackStopsFormation}.

\hl{Our simulations do not reach the mass resolution that has been achieved in some simulations in the literature} \citep[e.g.][]{motta2016}. \hl{As a consequence, we miss some star formation along the smaller-scale filaments. However, our 2m10k simulation does produce a power law IMF, like as expected from the literature} \citep[e.g.][]{Padoan_2007}. \hl{All our simulations have been done at the same numerical resolution. The trends we observe should hence be genuine.}

\hl{Various studies have been performed to investigate the limitations of simulations, particularly with regards to what physics are omitted and the impact that can have on results.}

\citep{outflowsIMF} \hl{looks at the effect of protostellar outflows on the IMF, they confirm that outflows limit the star formation rate.}

\hl{A series of papers by Lee and Hennebelle look at the effects of various initial conditions on the stellar mass spectrum and they find that all aspects of the IMF, the peak, low mass end, and high mass end, are sensitive to initial conditions} \citep[e.g.][]{lee_20181} and additional physics \citep[e.g.][]{Lee_20182,lee2019}.

We have performed SPH simulations of star cluster formation from molecular clouds with a polytropic index of unity for two cloud masses, varying only the mechanism for driving turbulence.

Clusters have been observed to have dynamical mass segregation \cite{massSegOBS} where more massive stars are preferentially found towards the centre of the cluster with the lower mass stars found further out. Our externally-driven runs show a trend consistent with this expectation. The details of the grouping are important though. As seen in Fig \ref{fig:radialmass}, while in the $2m10ke$ simulation at the chosen snapshot the envelope of the stellar masses smoothly declines with distance from the centre of mass, the $2m5ke$ run shows some substructure, where the distribution of masses also appear to be consistent with mass segregation.

\section{Summary and Conclusions}
In this work we directly compared star formation in externally driven clouds, akin the the GHC scenario \hl{and similar to clouds in some larger-scale simulations}, to clouds with turbulence \hl{driven everywhere, as in the gravoturbulent scenario}. We did this by comparing two driving methods, the first has turbulent driving across the entire box for the entire duration of the simulation, the other has turbulent driving prohibited in a central box within the simulated region. We compare various metrics between the two driving methods, as well as comparing to other simulation work and observations, including: velocity structure functions, IMF, mass segregation, cluster morphology, and cloud morphology.

\hl{All our simulations display a flow structure that is close to the expectations of supersonic turbulence, as evidenced by the velocity structure functions before we switch on gravity} (Fig. \ref{fig:structures}).

\hl{The edge-driven simulations develop a much more clustered mode of star formation} (Fig. \ref{fig:groups}). \hl{While we get power-law shapes for the IMFs in both cases} (Fig. \ref{fig:massfuncs}), \hl{the edge-driven simulations produce a lot more stars} (Fig. \ref{fig:SFE}) \hl{and show mass segregation in the formed groups} (Fig. \ref{fig:radialmass})


The morphology, both of the gas and the stellar population, shows a large difference between the two driving methods. The fully driven box has small scale clumps dispersed over the entire box with star formation spread out following the same pattern. The externally driven box shows a larger central clump surrounded with a system of filaments, this then produces a larger cluster in the centre of the clump with more stars forming along the filaments.

\hl{While the star formation details should not be compared to observations directly, due to the limited mass resolution in our simulations, this comparison shows that the differences in clustering properties between our gravoturbulent implementation (fully driven box) and GHC implementation (edge-driven box) conform to the expectations for the respective scenarios, i.e. the length scales on which a cloud is not supported by turbulence, but allowed to freely collapse, leaves an imprint in the size scale of the clustering of the stars.}

Star cluster formation with external driving turns out to be very similar to the turbulent collapsing cloud simulations \cite[e.g.][]{FeedbackStopsFormation,2014BateCollapse}, colliding flow simulations \citep[e.g.][]{collisionDObbs,collisionDobbs2,collidingmhdwurster}, and observations \citep[e.g.][]{massSegOBS} with star grouping and mass functions to be much more realistic than in our equilibrium turbulence simulations. \hl{Differences between our fully driven simulations and observations are likely due to limitations of our particular simulations (mass resolution in particular)}. Details of thermodynamics, \hl{MHD, and feedback} that we have not explored might affect this result \citep[e.g.][]{lee_20181,Lee_20182,outflowsIMF,grudic2018,Appel2022,Federrath_2015,padoan_2017}. However, from the simulations we have performed, we \hl{confirm} that the driving mechanism of molecular clouds has a huge effect on star formation.

\label{sec:conclusions}

\section*{Acknowledgements}

 SJ acknowledges support from the STFC grant ST/R00905/1. JDS acknowledges a studentship from the Science and Technology Facilities Council (STFC) (ST/T506126/1). We would like to thank the reviewer for their constructive criticism and helpful suggestions.

\section*{Data Availability}

Data and complete running instructions for all simulations are available on request: j.smith49@herts.ac.uk

Simulation movies will be available in the supplementary material on the journals website.
 



\bibliographystyle{mnras}
\bibliography{References} 





    


\bsp	
\label{lastpage}
\end{document}